\documentclass[pdflatex,sn-mathphys-num]{sn-jnl}

\usepackage{graphicx}%
\usepackage{multirow}%
\usepackage{amsmath,amssymb,amsfonts}%
\usepackage{amsthm}%
\usepackage{mathrsfs}%
\usepackage[title]{appendix}%
\usepackage{xcolor}%
\usepackage{textcomp}%
\usepackage{manyfoot}%
\usepackage{booktabs}%
\usepackage{algorithm}%
\usepackage{algorithmicx}%
\usepackage{algpseudocode}%
\usepackage{listings}%
\theoremstyle{thmstyleone}%

\theoremstyle{thmstyletwo}%

\theoremstyle{thmstylethree}%

\begin{document}

\title[Slow Quanta Bound States]{"Slow Quanta” Bound States and a Possible Link to Dark Matter}

\author*[1]{\fnm{Bruce} \sur{Denby}}\email{bruce.denby@sorbonne-universite.fr}

\affil*[1]{\orgdiv{Institut Langevin}, \orgname{ESPCI Paris, PSL University, CNRS, Sorbonne Universit\'e}, \orgaddress{\city{Paris}, \country{France}}}

\abstract{We study the possibility of elementary energy quanta with vacuum propagation speed \(w<c\), capable of interacting with each other to form massive bound states. The “slow matter” thus formed is shown to follow laws of Special Relativity mediated by velocity \textit{w} rather than \textit{c}, and to possess dynamical properties recalling some characteristics of Dark Matter.}


\maketitle

\section{Introduction}
\label{sec:intro}

Special Relativity is often introduced via \textit{gedanken} experiments in which light pulses are reflected from mirrors on fixed and mobile platforms, each equipped with a standard clock displaying the local time. The machinery of Special Relativity involves taking into account the round-trip travel times of light pulses sent to distant mirrors, the way in which these evolve as the platforms move, and the known speed of light, \textit{c}. 

Curiosity might lead one to ask whether relativistic effects will be observed if light is replaced by another type of wave, acoustic for instance, considering that target movement and pulse returns and are standard concerns in sonar and echolocation systems. An inspection of the acoustic Doppler shift formula, including Special Relativistic effects from source or target movement, gives interesting insight into this question. For an emitted sound frequency
\(f_0\), a velocity of approach \textit{v} between source and observer, and \textit{w} and \textit{c} the velocities of sound and light, we have \cite{Bachman1}\cite{Bachman2},

\begin{equation}
\label{eq:scrod1}
\begin{aligned}
f'_{ms} &= f_0\Bigl[\frac{w}{w-v}\Bigr]\times\sqrt{1-\frac{v^2}{c^2}}
\end{aligned}
\end{equation}

\begin{equation}
\label{eq:scrod2}
\begin{aligned}
f'_{mo} &= f_0\Bigl[\frac{w+v}{w}\Bigr]\div\sqrt{1-\frac{v^2}{c^2}}
\end{aligned}
\end{equation}

where \(f'_{ms}\) and \(f'_{mo}\) are the observed frequencies in the case of a moving source, or a moving observer, respectively. The bracketed terms in eq.~\ref{eq:scrod1} and~\ref{eq:scrod2} are precisely the \emph{non-relativistic} formulae for the acoustic Doppler shift, while the square root is a correction term due to relativistic time dilation, applied in the numerator for a moving source, or in the denominator for a moving observer. Since the result depends on the definition of fixed and mobile systems, the acoustic Doppler effect lacks relativistic invariance in the usual sense.

If we now consider \textit{w} to correspond to an \textit{arbitrary} wave disturbance, it is straightforward to see \cite{Bachman2} that choosing \(w=c\) causes the time dilation factors to exactly compensate the expressions in brackets, thereby merging eq.~\ref{eq:scrod1} and~\ref{eq:scrod2} into a single result~\ref{eq:scrod3} having no reference to a preferred frame. In fact, as shown in \cite{Bachman2}, \(w=c\) turns out to be the \emph{only} wave solution that has this property. Eq.~\ref{eq:scrod3} is, of course, the well known Doppler shift expression for light. 

\begin{equation}
\label{eq:scrod3}
\begin{aligned}
f' &= f_0\sqrt{\frac{1+\frac{v}{c}}{1-\frac{v}{c}}}
\end{aligned}
\end{equation}

In this situation, then, we do not obtain the equations of Special Relativity for sound. However, the observations made above allow us to make an interesting conjecture. Could the necessity for two equations in the acoustic Doppler formula be the result of a “mismatch” between the physical quantities being measured, that is, between a sound frequency in the bracketed expression, and a light-mediated time dilation mechanism in the square root term? Said another way, do we need two equations because acoustic phenomena are not disturbances in spacetime, but in ordinary matter known to undergo time dilation mediated by the speed of light \textit{c}, and not the speed of sound, \textit{w}?

If we entertain a positive reply, we may propose an alternate method of obtaining a single Doppler equation. Rather than taking \(w=c\) in eq.~\ref{eq:scrod1} and~\ref{eq:scrod2}, let us try \(c=w\). This clearly also works algebraically, but what is the physical meaning of such a choice? If we return to the case of acoustics, we are asked to imagine a kind of structure, or matter, made out of sound, which obeys a time-dilation formula with \textit{w}, the speed of sound, replacing \textit{c} as the limiting or invariant velocity. As it turns out, in the 1940's (see, e.g., \cite{Eshelby}), it was discovered that plastic deformations in certain materials give rise to acoustic “sine Gordon” solitons, whose pulse widths and effective masses obey a version of relativity mediated by the velocity of sound in the medium. The result generated considerable interest in the literature (e.g. \cite{Bergman}), the interpretation being that the inner dynamics of the soliton disturbances, that is, their time dilation factor, is determined by the velocity of \textit{phonons} rather than of photons. Other examples of acoustic analogs of relativity can be found in \cite{Unruh 1981}\cite{Unruh 2014}\cite{Gordon 1923}.

In what follows, we extend the idea of an “alternate relativity” to spacetime itself, by asking the question, could there exist in the universe energetic disturbances, akin to light but propagating at a speed \(w < c\), capable of forming matter that obeys Special Relativity mediated by \textit{w} rather than \textit{c}? Exploring this question, and its implications, is the objective of the present article.

In the next section, we outline the hypotheses and definitions forming the basis of the ensuing analyses, and introduce a “slow matter” bound state displaying the necessary time dilation properties. In section~\ref{sec:lorentz}, we confirm that a complete Lorentz transformation built around waves of velocity \(w<c\) can indeed be constructed for a “slow matter sector”. Section~\ref{sec:dynamic} describes the dynamic properties of slow matter and its interactions with normal matter, and presents ways in which the latter bear interesting similarities to interactions of dark matter with normal matter. Conclusions and some propositions for further investigations appear in the final section.

\section{Hypotheses and Definitions}
\label{sec:hypotheses}

Wave phenomena, arising from an abundance of underlying physical phenomena, are ubiquitous in nature. In a given medium, it also sometimes occurs that there exist multiple types of waves having different propagation velocities. Some examples are shear waves in ultrasonics, with velocities some one thousand times less than standard longitudinal waves; second sound disturbances in superfluids, propagating at about one-tenth the velocity of regular sound waves; and the complicated dispersion characteristics of seismic waves depending upon their type. Our starting hypothesis is that spacetime, as well, could turn out to harbor some as yet undiscovered elementary sub-light-speed waves. 

\subsection{Time Dilation and Light Clocks}

Some aspects of Special Relativity amount to simple corrections for the speed of the wave used to make measurements, as are also found in acoustic echolocation systems and the like. The phenomenon of time dilation, however, is special in that it appears to involve the flow of energy within matter itself, rather than in light beams emitted externally to perform \textit{gedanken} experiments. The behavior of time dilation is often presented in analogy to a “light clock”, consisting of a light beam bouncing back and forth between two mirrors, separated by a fixed distance, contained within a moving system. As an example, if we orient such a clock with its light beam perpendicular to the system’s movement, it is easy to see that the light will take a diagonal path rather than a straight one as the system moves, making the clock’s oscillation appear, to an outside observer, to have effectively slowed down – the phenomenon of time dilation. 

Although ordinary matter does not contain mirrors and so on \textit{per se}, the mechanism of time dilation in particles, atoms, etc., is believed to arise from the identical principle. Physicists accept that the time dilation factor of Special Relativity, which is identical by construction to that of the light clock, is correct, since no deviation from Special Relativity has ever been detected. The light clock, furthermore, has meanwhile become something of a standard tool in both theoretical and experimental studies in relativity, e.g. \cite{Sfarti 2018}\cite{Bravo 2023}\cite{Ya’acov 2023}. In the present work, we propose to study an alternate time dilation mediated by waves traveling at velocity \textit{w}, by way of a system modeled on a light clock design, as described in the next section.

\subsection{A Slow Matter Bound State}

\begin{figure}[htbp]
\centering
\includegraphics[width=.4\textwidth]{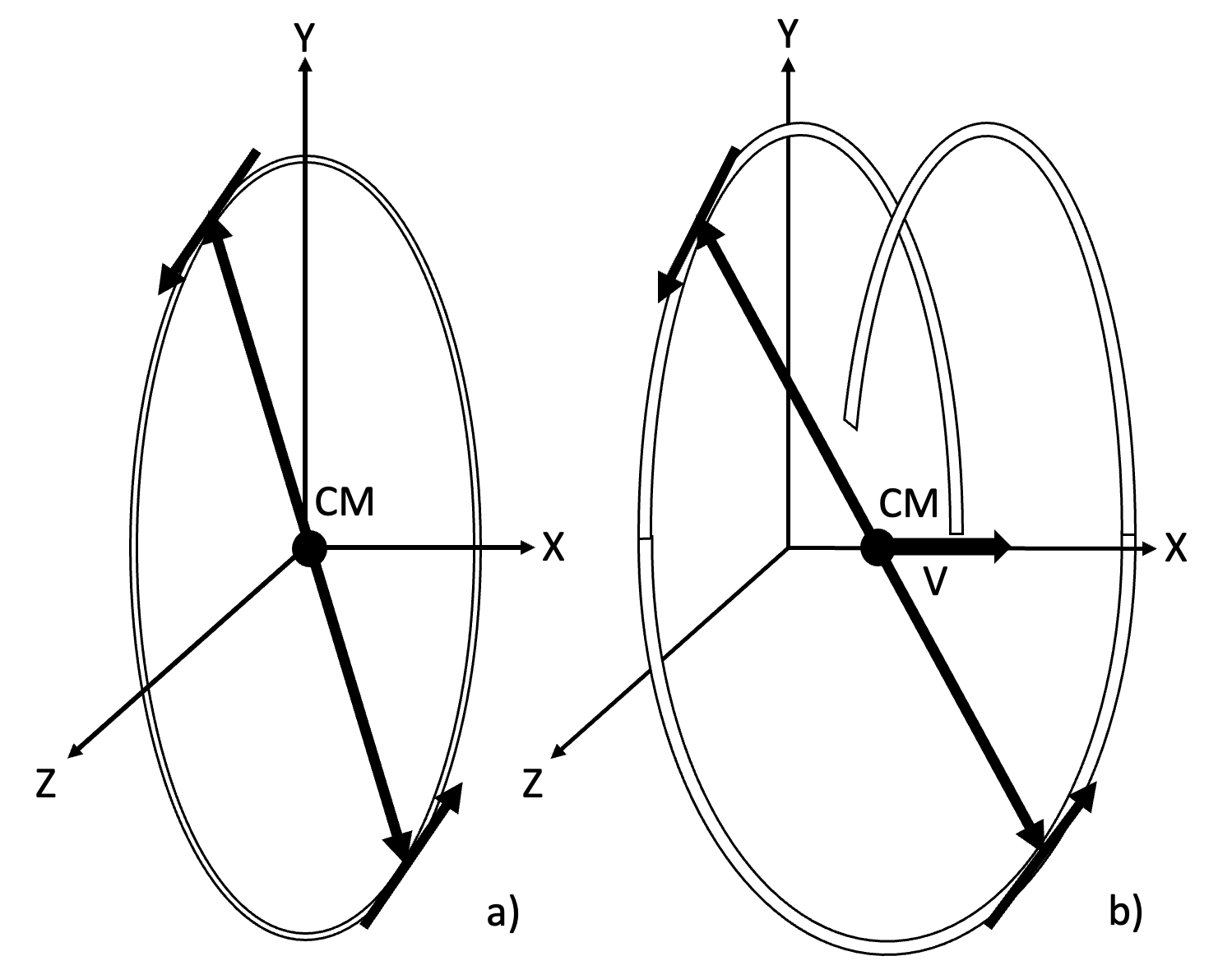}
\caption{Slow matter bound state comprised of two self-orbiting slow quanta (arrows), showing the center of mass, CM, at rest, a), and boosted to velocity \textit{v}, b).}
\label{fig:boundstate}
\end{figure}

We posit the existence of “slow” elementary energy quanta of vacuum propagation velocity \(w<c\) that interact between themselves via an unspecified attractive force, allowing them to form stable bound states. While the slow quanta themselves do not interact with other forms of matter, the bound states, having mass, can interact with other massive bodies via gravitation. Fig.~\ref{fig:boundstate} illustrates a bound state composed of two slow quanta orbiting each other in a circle, shown at rest on the left, and boosted to a constant nonzero velocity, on the right.

It is evident from fig.~\ref{fig:boundstate} that the path of the slow quanta in the boosted bound state is a helix, which, when unrolled into two dimensions, yields a diagonal path for the slow quanta that is identical to that taken by the beam of a light clock. The bound state therefore patently undergoes \textit{w}-mediated time dilation when moving at a fixed velocity.   

There are several possibilities for the force holding the bound state together. \textit{Geons} are self-supporting structures of light \cite{Wheeler 1955}, gravity waves \cite{Brill and Hartle 1964}, or gravitons \cite{Guiot et al. 2020}, that have been studied in the literature and sometimes \cite{Guiot et al. 2020} proposed as Dark Matter candidates. Geons, however, are composed of ordinary \textit{c}-mediated matter, and are generally portrayed as very massive bodies. More plausible for our bound state would be a separate mediator field binding the slow quanta together; or simpler, self-interaction between the quanta, which is known to lead to bound states in modified sine Gordon models \cite{Delfino}. As a knowledge of the binding mechanism is not necessary in what follows, it will be left unspecified.

To set the stage for the discussion of Lorentz transformations in the next section, we suppose that the bound state can emit and absorb slow quanta, of negligible energy compared to its mass so as to leave its kinematics unperturbed, in order to provide a mechanism for measuring the coordinates of spacetime events. Also, since, other than gravity, there is no interaction between slow matter and normal matter, we assume that measurements between the two types of systems take place through a type of gravitational lensing - of light, on the one hand, and of slow quanta, on the other.

\section{A Lorentz Transformation for Slow Matter}
\label{sec:lorentz}

This section develops a new Lorentz transformation governing the motion of slow matter bound states. Being mediated by \textit{w} rather than \textit{c}, the bound states manifest a departure from Poincaré symmetry in the usual sense. This somewhat damning realization is however mitigated by a few observations. The first is that, although no violation of Poincaré symmetry has ever been detected, tests of it continue to this day, as for example in \cite{CHOOZ 2012}, \cite{IceCube 2014} and \cite{D0 2015}. It is also the case that the search for an explanation of Dark Matter has already assembled a wide range of sometimes outlandish possibilities, including particles such as tachyons that also do not obey Poincaré symmetry \cite{New Era} \cite{tachyons 2024}. Finally, in this work, slow matter and normal matter are assumed not to interact with each other except via gravitational lensing, so that a violation of Poincaré symmetry would be difficult to formally detect.

\subsection{Deriving the Lorentz Transform}

In what follows, we use the previously described properties of slow matter and simple vector algebra to develop a functional Lorentz transformation for matter mediated by quanta of velocity \(w<c\). The procedure is fairly simple; however, the derivation of a Lorentz transformation – even a standard one – from first principles and vector algebra, without recourse to the Einstein postulates, is not something that is widely discussed in the literature. Therefore, for clarity, we outline the procedure here. 

The treatment will be carried out in one spatial dimension \textit{x} and the time dimension \textit{t} using standard Minkowski or \textit{x-t} diagrams, where the direction of motion of moving systems is also in the \textit{x} direction. The extension to three spatial dimensions may be carried out using traditional methods. A system is assumed able to send out wavefronts towards, and receive reflections back from, external spacetime events, in order to make measurements. Each system also contains a clock whose rate is governed by the properties of a light clock employing quanta of the appropriate mediating velocity, for example, \textit{w}.

\begin{figure}[htbp]
\centering
\includegraphics[width=.4\textwidth]{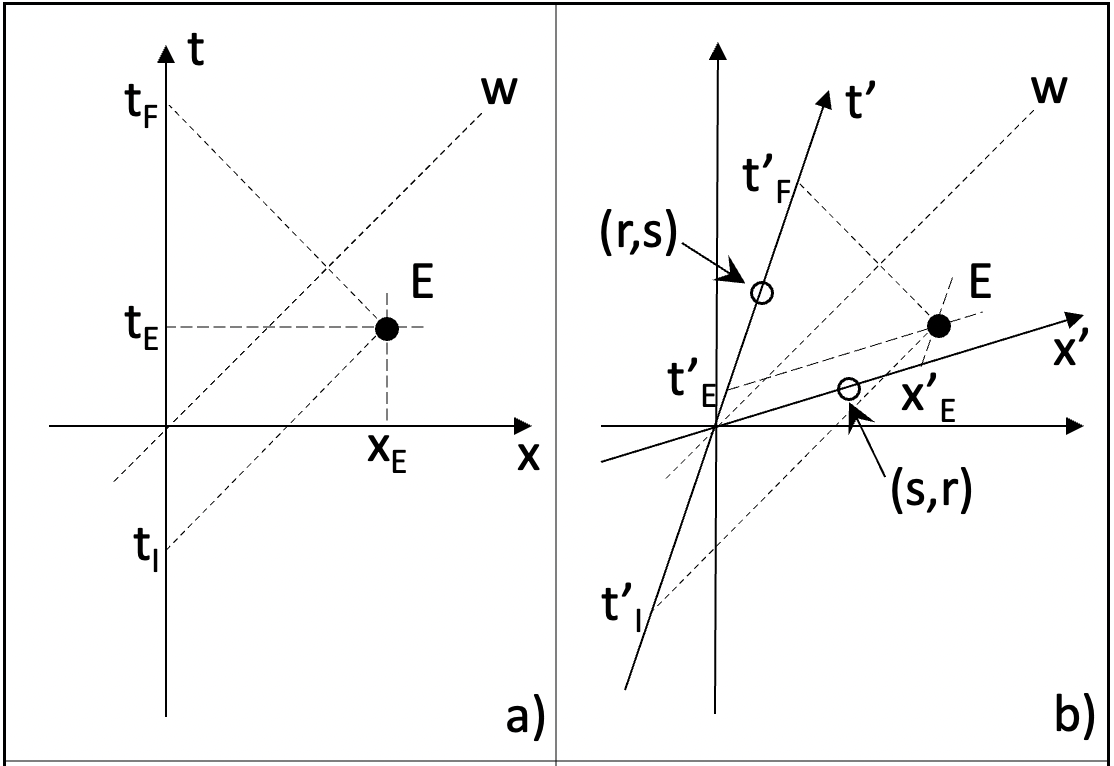}
\caption{In a) the fixed, unprimed system emits a quantum at \(t_I\) to measure the coordinates \((x_E,t_E)\) of event E. In b), the same event E is measured as \((x’_E,t’_E)\) in the moving system, by emitting a quantum at time \(t’_I\). Paths of emitted quanta are shown as dotted lines of unit slope, while measurements appear as dashed lines. The points (s,r) and (r,s) are discussed in the text.}
\label{fig:lorentz}
\end{figure}

Fig.~\ref{fig:lorentz}a shows an unprimed, fixed system with axes \((x,t)\), and fig.~\ref{fig:lorentz}b a primed system moving at velocity \textit{v} in the \textit{x} direction and having axes \((x’,t’)\). Both systems measure the spacetime coordinates of an event indicated by point E. We note that the time axis of the moving system has slope \(1/v\) with respect to the vertical. For simplicity, we have also normalized the speed of the quanta, \textit{w}, to 1 here, so that lines representing quanta have slope 1, as indicated by the lines “\textit{w}”. To measure the coordinates of the event E, the unprimed system emits a quantum at time \(t_I\) in the (1,1) direction, which is then reflected by event E and returned along the (-1,1) direction to the origin, arriving at time \(t_F\). As the unprimed system is not moving, the outgoing and return times are the equal, so we may write:

\begin{equation}
\label{eq:scrod4}
\begin{aligned}
x_E &= \frac{t_F-t_I}{2}
\end{aligned}
\end{equation}

\begin{equation}
\label{eq:scrod5}
\begin{aligned}
t_E &= \frac{t_F+t_I}{2}
\end{aligned}
\end{equation}

Because the primed system is moving, when it sends and receives its quanta, the corresponding times \(t'_I\) and \(t'_F\) will not be symmetric, as shown in the figure. Nevertheless, since the system is unaware of its movement, it uses the same method as the unprimed system to obtain the coordinates of E:

\begin{equation}
\label{eq:scrod4a}
\begin{aligned}
x'_E &= \frac{t'_F-t'_I}{2}
\end{aligned}
\end{equation}

\begin{equation}
\label{eq:scrod5a}
\begin{aligned}
t'_E &= \frac{t'_F+t'_I}{2}
\end{aligned}
\end{equation}

Let the unprimed coordinates \((r,s)\) label the point corresponding to one unit of time along the moving frame’s time axis \(t'\), as indicated in fig.~\ref{fig:lorentz}b by a small open circle. With \(\lambda\) and \(\mu\) as arbitrary multipliers, we may write:

\begin{equation}
\label{eq:scrod6}
\begin{aligned}
(x_E,t_E)-\lambda(1,1) &= t'_I(r,s)
\end{aligned}
\end{equation}

\begin{equation}
\label{eq:scrod7}
\begin{aligned}
(x_E,t_E)+\mu(-1,1) &= t'_F(r,s)
\end{aligned}
\end{equation}

The multipliers \(\lambda\) and \(\mu\) are easily eliminated to yield:

\begin{equation}
\label{eq:scrod8}
\begin{aligned}
t'_I &= \frac{(x_E-t_E)}{(r-s)}
\end{aligned}
\end{equation}

\begin{equation}
\label{eq:scrod9}
\begin{aligned}
t'_F &= \frac{(x_E+t_E)}{(r+s)}
\end{aligned}
\end{equation}

Substituting ~\ref{eq:scrod8} and ~\ref{eq:scrod9} into ~\ref{eq:scrod4a} and ~\ref{eq:scrod5a} above, we find:

\begin{equation}
\label{eq:scrod10}
\begin{aligned}
x' &= \frac{(-sx+rt)}{(r^2-s^2)}
\end{aligned}
\end{equation}

\begin{equation}
\label{eq:scrod11}
\begin{aligned}
t' &= \frac{(rx-st)}{(r^2-s^2)}
\end{aligned}
\end{equation}

where we have dropped the subscript E since an event E can be at any point in the plane. In matrix form, the transformation and its inverse become: 

\begin{equation}
\label{eq:scrod12}
\begin{aligned}
\begin{pmatrix} x'\\t'
\end{pmatrix}
&=\frac{1}{(r^2-s^2)}
\begin{pmatrix} -s&r\\r&-s
\end{pmatrix}
\begin{pmatrix} x\\t
\end{pmatrix}
\end{aligned}
\end{equation}

\begin{equation}
\label{eq:scrod13}
\begin{aligned}
\begin{pmatrix} x\\t
\end{pmatrix}
&=
\begin{pmatrix} s&r\\r&s
\end{pmatrix}
\begin{pmatrix} x'\\t'
\end{pmatrix}
\end{aligned}
\end{equation}

It is easy to verify that the unprimed system vectors \((r,s)\) and \((s,r)\), shown as small open circles in Fig.~\ref{fig:lorentz}b, correspond to the vectors (0,1) and (1,0) in the primed system, that is, to unit vectors along, respectively, the \(t’\) and \(x’\) axes. Since linear transformations are area-preserving, the area of the parallelogram defined by \((r,s)\) and \((s,r)\) must be the same as that of the square defined by the points (0,1) and (1,0) in the unprimed system, or, in other words, \((s^2 - r^2) = 1\), so that the normalization factor in ~\ref{eq:scrod12} can be replaced by -1. Re-inserting explicitly the velocity of the primed system, \textit{v}, and of the quanta, \textit{w}, and defining \(\gamma_w\) as,

\begin{equation}
\label{eq:scrod14}
\begin{aligned}
\gamma_w &= \frac{1}{\sqrt{1-\frac{v^2}{w^2}}}
\end{aligned}
\end{equation}

we obtain for the transform and its inverse:

\begin{equation}
\label{eq:scrod15}
\begin{aligned}
\begin{pmatrix} x'\\t'
\end{pmatrix}
&=\gamma_w
\begin{pmatrix} 1&-v\\-v/w^2&1
\end{pmatrix}
\begin{pmatrix} x\\t
\end{pmatrix}
\end{aligned}
\end{equation}

\begin{equation}
\label{eq:scrod16}
\begin{aligned}
\begin{pmatrix} x\\t
\end{pmatrix}
&=\gamma_w
\begin{pmatrix} 1&v\\v/w^2&1
\end{pmatrix}
\begin{pmatrix} x'\\t'
\end{pmatrix}
\end{aligned}
\end{equation}

This completes the derivation of a Lorentz transform that, provided the starting hypotheses are verified, is valid for an arbitrary value, \textit{w}, of the velocity of the quantum assumed to mediate the properties of matter. We note, finally, that for systems possessing a single, fixed value of \textit{w}, the transform still displays a “Poincaré-like” symmetry; whereas when comparing \textit{w}-systems and \textit{c}-systems, Poincaré symmetry in the usual sense is not upheld.  

\subsection{Slow Matter and Normal Matter Together}

\begin{figure}[htbp]
\centering
\includegraphics[width=.4\textwidth]{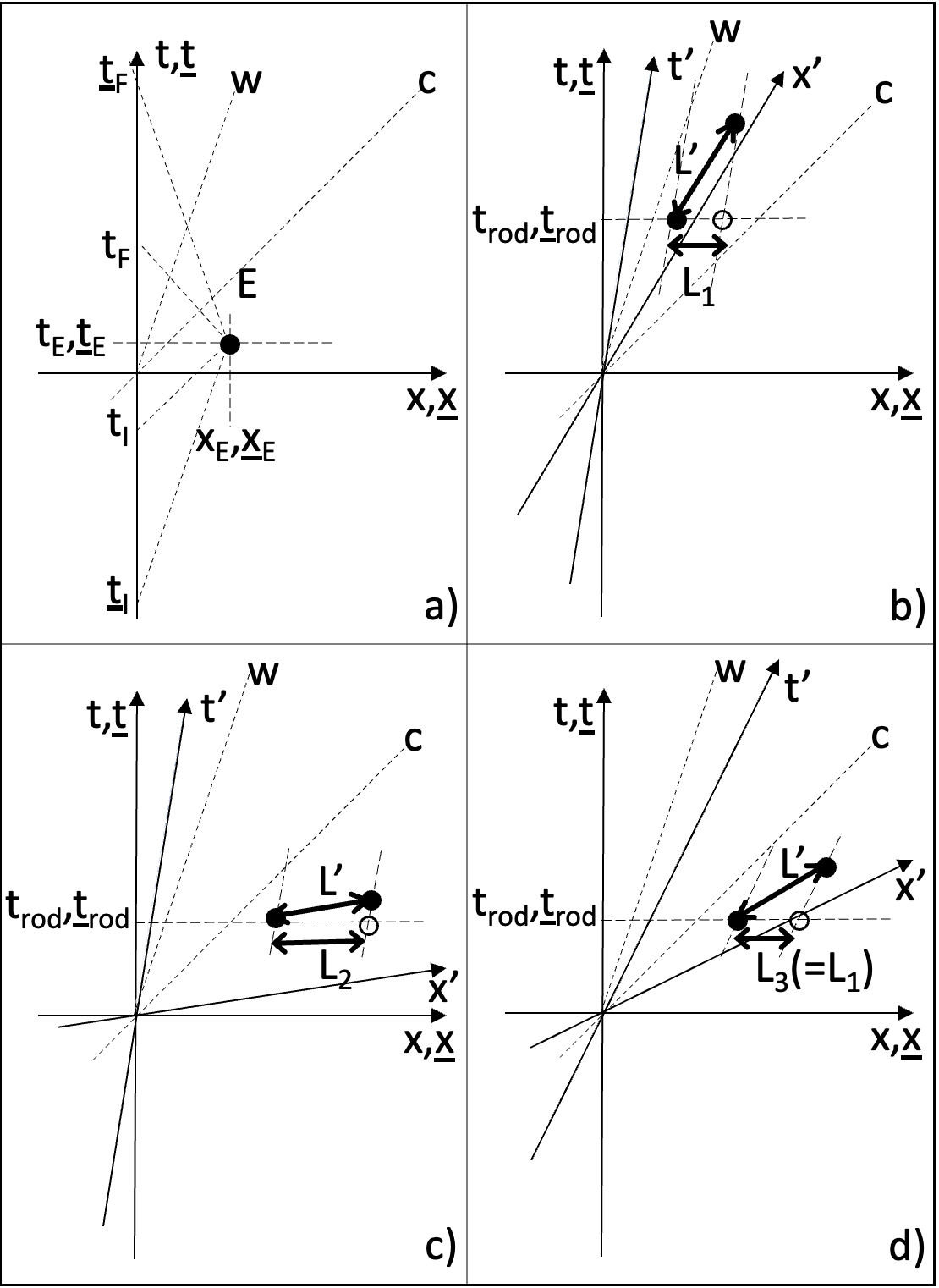}
\caption{In~\ref{fig:together}a), the spacetime event E is localized by a normal matter system \((x,t)\) and a slow matter system \((\underline{x},\underline{t})\). In~\ref{fig:together}b), the \((x,t)\) and \((\underline{x},\underline{t})\) systems measure the length of a slow matter rod of proper length \(L’\) in the system \((x’,t’)\) moving at velocity \textit{v}. In~\ref{fig:together}c), the \((x,t)\) and \((\underline{x},\underline{t})\) systems measure the length of a normal matter rod of proper length \(L’\) in the system \((x’,t’)\) moving at velocity \textit{v}. In~\ref{fig:together}d), the \((x,t)\) and \((\underline{x},\underline{t})\) systems measure the length of a normal matter rod of proper length \(L’\) in the system \((x’,t’)\) moving at velocity \(v_{new}\) where \(v_{new}/c = v/w\) and \(v_{new} > w\). The measured lengths \(L_1\), \(L_2\) and \(L_3\) (\(=L_1\)) are discussed in the text.}
\label{fig:together}
\end{figure}

In the following sections, we will study interactions between slow matter and regular matter systems. For this, it is necessary to display the spacetime transformation properties of both systems on the same plot. Some illustrative examples are shown in fig.~\ref{fig:together}.

In fig.~\ref{fig:together}a, the coordinate axes of an unprimed normal matter system \((x,t)\), and an underlined slow matter system  \((\underline{x},\underline{t})\), both at rest, are superimposed to show how these systems measure the spacetime event E in this situation. Velocities are normalized to \(c = 1\), indicated by the 45 degree dotted line labeled \textit{c}. Paths of slow quanta emitted with velocity \(w<c\) will thus be parallel to the dotted line labeled \textit{w}. In fig.~\ref{fig:together}a), both systems emit quanta, shown by dotted lines, which are reflected by the spacetime event E back to their respective origins. Since the unprimed and underlined systems are both at rest, and both use the method of equations~\ref{eq:scrod4} and~\ref{eq:scrod5} to make their measurements, it is evident from fig.~\ref{fig:together}a that the times measured by the slow matter and normal matter system for event E will be identical, provided the systems use the same time base when at rest. It is also clear that the distances measured by the unprimed and underlined systems will be identical as well, provided we scale these by the velocity of the quantum used to make the measurement, i.e., by \textit{c} in the case of light, and by \textit{w} in the case of slow matter.
  
In fig.~\ref{fig:together}b we introduce a third system, again of slow matter, with coordinates \((x’,t’)\), moving at velocity \textit{v} with respect to the unprimed (normal matter) and the underscored (slow matter) systems. The time axis, labeled \(t’\), and space axis \(x’\) appear asymmetrically about the line “\textit{w}”, due to the orientation of the line “\textit{w}” as compared to the line “\textit{c}”. We concern ourselves here with two spacetime events, corresponding to the endpoints of a rigid rod of proper length \(L’\) in the primed slow matter system, that appears as a double-headed arrow parallel to the \(x’\) axis in the figure. The world lines of the two endpoints appear as short dashed lines parallel to the \(t’\) axis. To obtain the length of the rod as measured in the fixed, underlined slow matter system, we apply the Lorentz transformation of eqs.~\ref{eq:scrod15} and~\ref{eq:scrod16} connecting the two systems. Due to length contraction, we obtain, in the fixed underlined slow matter system, the length \(L_1 = L’/\gamma_w\) indicated by a shorter double-headed arrow parallel to the \(x\) and \(\underline{x}\) axes. In the unprimed, fixed normal matter system, the length \(L_1\) is also obtained, since, as demonstrated in the discussion of fig.~\ref{fig:together}a, systems at rest always obtain the same spacetime coordinates for an event. The length measurements are carried out at time \(t_{rod} = \underline{t}_{rod}\) as shown in the figure. Thus, a moving slow matter rod of proper length \(L’\) will appear to have a length \(L_1 = L’/\gamma_w\) in a fixed slow matter system, as well as in a fixed normal matter system.
  
In fig.~\ref{fig:together}c, we repeat the experiment of fig.~\ref{fig:together}b, but this time the moving system, still of velocity \textit{v}, contains a normal matter rod of proper length \(L’\) rather than a slow matter rod. A standard \textit{c}-mediated Lorentz transformation, articulated about the line “\textit{c}”, is applied. This time, a rod length \(L_2 = L’/\gamma_c > L_1\), where \(\gamma_c\) simply replaces \textit{w} by \textit{c}, is obtained, both for the fixed regular matter system and the fixed slow matter system. Thus, although the relative velocity between the slow matter system and the regular matter system are identical in fig.~\ref{fig:together}b and fig.~\ref{fig:together}c, the measured length of the moving rod is different in the two systems, demonstrating Poincaré non-invariance when comparing slow and normal matter systems.

In the final example, fig.~\ref{fig:together}d, we repeat the experiment of fig.~\ref{fig:together}c but choose the normal matter system velocity \(v_{new}\) so that \(v_{new}/c = v/w\) and \(v_{new} > w\). Defining a \(\gamma_{c,new}\) that incorporates \(v_{new}\) rather than \textit{v}, the measured length of the moving rod is now \(L_3 = L’/\gamma_{c,new}\), both in the fixed normal matter system and in the fixed slow matter system – despite the fact that the rod’s movement is “superluminal”, i.e., exceeds \textit{w}, in the slow matter system. We also remark that since \(\gamma_{c,new} = \gamma_{w} \) in this example, we have \(L_3 = L_1\) as in fig.~\ref{fig:together}b.

We have seen that a measured property of a moving reference frame, due to the different time dilation mechanisms, depends upon the matter type making up the moving frame. Thus, to make a measurement on a moving system of matter type A, as an example, a type B observer must first transform to a hypothetical fixed type A system. Transformations between arbitrary slow and regular matter systems can then be obtained via a succession of homogeneous transforms between moving and fixed systems of the same matter type. The examples shown above demonstrate that relativistic effects in slow matter will indeed be observed in normal matter systems. The situation is analogous to the laboratory measurements of relativistic acoustic solitons mentioned earlier \cite{Eshelby}\cite{Bergman}, in that both laboratory and acoustic systems are fixed, and the time dilation factor observed for the moving soliton is that of the acoustic system, not of the laboratory system. The same time dilation factor, furthermore, was also found to intervene in the measured effective mass of the solitons. In the next section, we turn our attention to the dynamic properties of slow matter.

\section{Dynamic Properties of Slow Matter}
\label{sec:dynamic}

This section presents a first look at a scenario in which bulk slow matter and normal matter exist in the universe and can interact dynamically through gravity on a cosmological scale. Note that we assume slow mass and normal mass both gravitate via the usual universal constant, G. We begin by establishing the momentum-space behavior of slow matter.

\subsection{The Slow Matter Momentum-Space 4-Vector}

It is well known (see for example \cite{FeynmanLectures}) that the Lorentz transformation for momentum-space 4-vectors can be obtained from the formula for relativistic composition of velocities, along with three assumptions: that the momentum \(p = mv\), with \textit{m} the total mass and \textit{p} and \textit{v} vectors pointing along the same direction; that momentum is conserved; and that the total mass of an object is a function of its velocity \textit{v}. If we accept the premise that slow matter possesses mass, energy, and momentum of the same character as the analogous quantities in normal matter, it is simple, beginning with eqs.~\ref{eq:scrod15} and~\ref{eq:scrod16}, to use the same procedure to obtain an energy-momentum 4-vector for slow matter, with the end result that one simply replaces \textit{c} by \textit{w} in the standard definition\footnote{We assume that the energy, momentum, and mass of slow matter are like those of normal matter. One might expect, however, that the relation E = \(h\nu\), with \textit{h} Planck’s constant and \(\nu\) the frequency of an energy quantum, would not hold, since there would then be an ambiguity between normal matter and slow matter.}.

Making use of the Lorentz formalism derived in sec.~\ref{sec:lorentz}, now extended to three spatial dimensions, we verify that applying a slow matter boost to a slow matter bound state 4-vector is equivalent to boosting the 4-vectors of its constituent self-orbiting quanta. Referring to fig.~\ref{fig:boundstate}, we consider an instant when the two quanta making up the bound state have momenta \(p_y\) and \(-p_y\) oriented in the \(+y\) and \(-y\) directions respectively. Taking the energy of each quantum as \(E\), their initial 4-vectors will be given by \((E, 0, p_y, 0)\) and \((E, 0, -p_y, 0\)). Applying a boost of velocity \textit{v} in the \(+x\) direction to these, we obtain the boosted 4-vectors \((E\gamma_w, \gamma_wvE/w^2, p_y, 0)\) and \((E\gamma_w, \gamma_wvE/w^2, -p_y, 0)\). The sum of these two vectors, that is, the boosted 4-vector of the entire slow matter bound state, is given by \((2E\gamma_w, 2E\gamma_wv/w^2, 0, 0)\). Interpreting \(2E/w^2\) as the rest mass \(m_0\) of the bound state, we obtain for the energy of the boosted bound state \(E_{boosted} = (m_0\gamma_w)w^2\), and for its momentum \(p_{boosted} = (m_0\gamma_w)v\). These are the standard definitions of the energy and momentum of a body of rest mass \(m_0\) boosted to a velocity \textit{v}, transposed here to a system where Special Relativity is mediated by the velocity of slow quanta \textit{w} rather than by the speed of light \textit{c}.

\subsection{Two Body Mechanics of a “Toy” Slow Matter-Regular Matter Interaction}

If the intrinsic velocity \textit{w} mediating the behavior of slow matter is considerably lower than \textit{c}, it might be expected that gravitational interactions between comparable amounts of slow matter and regular matter would lead to highly relativistic instances of slow matter. These slow matter “bodies” would in turn gain relativistic mass, potentially leading them to play a dominant role in the gravitational dynamics of a normal/slow matter system.

The proper formalism for exploring such an idea would undoubtedly be General Relativity, however the question of how exactly to incorporate “slow matter” into General Relativity suggests a more heuristic approach to obtain a first idea. The situation is similar to using the special-relativistic mass increase of Mercury to estimate the precession of the planet’s perihelion, which has been shown to give results similar in form and magnitude to the full General Relativity solution \cite{Roseveare 1982}\cite{Lemmon and Mondragon 2016}. Indeed, the justifications advanced in the perihelion studies, e.g., non-radiative case, absence of singularities, and so on, apply also in the slow matter case at hand. An exception, of course, is that while Mercury’s precession is a small perturbation to its orbit, for slow matter, it is the highly relativistic regime that is likely to be of interest. Nonetheless, if the velocity \textit{w} is indeed well below \textit{c}, relativistic effects should be important only for the slow sector, not for the normal matter component.

\begin{figure}[htbp]
\centering
\includegraphics[width=.4\textwidth]{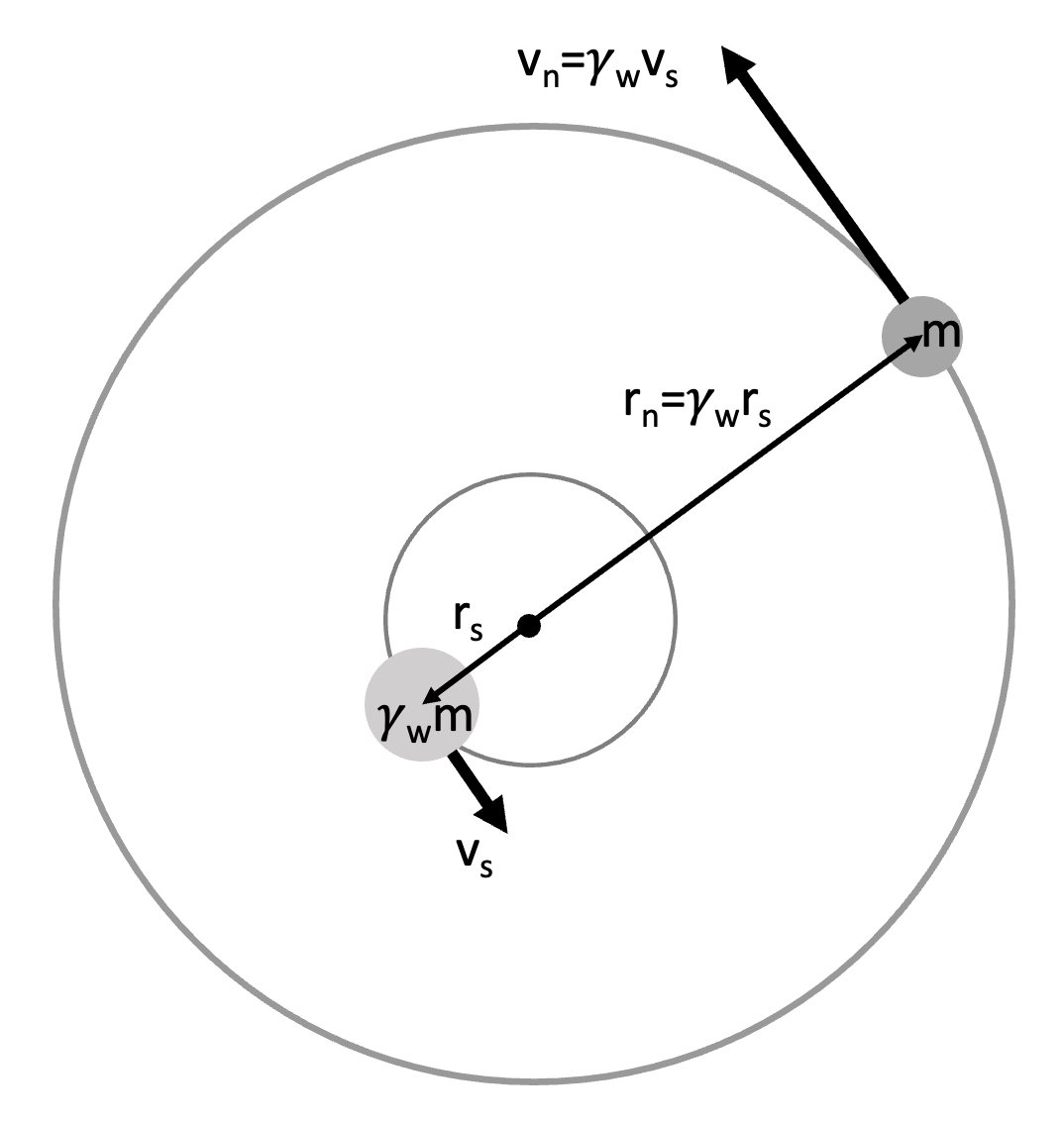}
\caption{A circular binary system consisting of two bodies of equal rest mass \textit{m}, one made of slow matter, with orbital radius \(r_s\) and velocity \(v_s\), and the other of normal matter, having orbital radius \(r_n\) and velocity \(v_n\). In the analysis proposed in the text, as the slow matter body gains velocity, its mass increases by the factor \(\gamma_w\), leading to radii and velocities that also scale by this factor, as indicated in the figure.}
\label{fig:celestial}
\end{figure}

With these caveats, we present a “toy” system consisting of two bodies of equal mass, \textit{m}, one composed of slow matter, and the other of normal matter, that interact gravitationally. Taking \(v_n\) and \(v_s\) as the velocities of the normal and slow matter bodies, respectively, and if the center of mass of the system is at rest, we will have, by conservation of momentum, that \(\gamma_wmv_s = mv_n\), or equivalently, \(v_s = v_n/\gamma_w\). Thus, as \(\gamma_w\) increases, the slow matter body becomes the heavier object and begins to play a structuring role in the orbit obtained. We will assume for simplicity a circular binary system as in fig.~\ref{fig:celestial}, so that we also have for the radii of the orbits, \(r_n = \gamma_wr_s\). Equating the attractive gravitational force between the bodies to the centrifugal force necessary to retain the slow body in its orbit, taking \(r = r_n+r_s\), and assuming \(\gamma_w\) large, we obtain:

\begin{equation}
\label{eq:scrod17}
\begin{aligned}
\frac{\gamma_wm^2G}{r^2} &= \frac{\gamma_w m{v_s}^2}{r_s}
\end{aligned}
\end{equation}

\begin{equation}
\label{eq:scrod18}
\begin{aligned}
\frac{mG}{r} &= (\gamma_w+1){v_s}^2 
\end{aligned}
\end{equation}

\begin{equation}
\label{eq:scrod19}
\begin{aligned}
\frac{mG}{r} \approx \gamma_w{v_s}^2 = (\gamma_w v_s)v_s
\end{aligned}
\end{equation}

\begin{equation}
\label{eq:scrod20}
\begin{aligned}
\frac{mG}{r} \approx (v_n)v_s \equiv (\gamma_w w)w 
\end{aligned}
\end{equation}
 
In the last step, we have considered that when \(\gamma_w\) is large, \(v_s\) is near \textit{w}, with \(v_n\) being determined by the actual value that \(\gamma_w\) takes on. Thus, we consider that the slow matter body is circling the center of mass at a low velocity near \textit{w}, while the the normal matter body, located \(\gamma_w\) times further away from the center of mass, does so at a velocity that is \(\gamma_w\) times greater. Thus, without making a statement on the genesis or stability of the result, we note that the dynamics of slow matter in this simple example play a structuring role in the character of the orbit obtained. This unusual characteristic of slow matter kinematics can be considered a means of storing kinetic energy locally in the form of relativistic mass.        

\subsection{Possible Parallels to Dark Matter}

Dark Matter makes up 80 percent of the matter in the universe, yet despite almost 100 years of theoretical conjectures and experimental tests, its composition remains unknown. In current models of cosmic evolution, Dark Matter sub-halos were first to develop, creating potential wells where normal matter could accumulate to form stars and finally larger structures. 

Dark Matter furthermore principally interacts with normal matter through gravitation, and is studied either via lensing \cite{lensing} or so-called Direct Detection techniques \cite{direct detection} where models of Dark Matter interactions can be tested on signals produced when terrestrial detectors pass though regions of our galaxy containing Dark Matter. For the latter method, a knowledge of Dark Matter velocities is important. Though galactocentric velocities of Dark Matter particles are difficult to characterize, in 2014 \cite{54mps}, theoretical considerations estimated a mean velocity of Dark Matter particles of only 54 meters per second; while in 2019 \cite{metal poor}, data from the Gaia satellite suggested a mean velocity consistent with zero for metal-poor halo stars co-moving with Dark Matter sub-halos.

The surprisingly low estimates of Dark Matter velocities and the structuring nature of Dark Matter sub-halos suggest parallels to the properties evoked here for slow matter. Indeed, some researchers have suggested that as more and more explanations are ruled out or lead to dead ends, the Dark Matter search has entered a New Era \cite{New Era} where exotic possibilities like tachyons \cite{tachyons 2024} or other non-Poincaré constructs should be considered in a more serious way. Perhaps slow matter also falls into this category.

\section{Conclusions and Perspectives}
\label{sec:conclusion}

Assuming the existence of “slow” elementary energy quanta capable of forming massive bound states, we have shown that a consistent Lorentz framework can be constructed that predicts novel kinematic properties for slow matter, some of which recall properties of Dark Matter. The unusual kinematics predicted for slow matter constitute a means of storing kinetic energy locally in the form of relativistic mass. From a theoretical standpoint, it would be interesting to explore the current results in light of a more rigorous General Relativity based study of slow matter. On the experimental side, if a new special limiting velocity \textit{w} indeed exists, one might expect it to manifest itself in some way in the kinematics of Dark/Normal matter structures accessible to earthbound scientists and the instruments they exploit. Finally, we remark the Lorentz framework presented in the paper is formally valid for \(w > c\) as well, which could be another interesting case to explore.

\backmatter

\bmhead{Acknowledgements}

The author acknowledges fruitful exchanges with Steven J. Yellin, Norman Wildberger, William G. Scott, Lenny Siegel, and Jeffery Wyss.

AI tools were used to help locate and organize relevant background material, but not at all in the conception or preparation of the manuscript. 

\section*{Declarations}

\bmhead{Funding}

The author did not receive support from any organization for the submitted work.

\bmhead{Competing interests}

The author declares that they have no competing interests.

\bmhead{Ethics approval and consent to participate}

Not applicable.

\bmhead{Consent for publication}

Not applicable.

\bmhead{Data availability}

No datasets were generated or analyzed in the study.

\bmhead{Materials availability}

Not applicable.

\bmhead{Code availability}

Not applicable.

\bmhead{Author contributions}

The author carried out the conception and development of the work and drafting and revising the manuscript. The author has read and approved the final manuscript.


\begin{thebibliography}{99}

\bibitem{Bachman1}
R.A. Bachman,
\emph{Relativistic Acoustic Doppler Effect},
\emph{ Am. J. Phys.} {\bf 50} (1982) pg. 816-818

\bibitem{Bachman2}
R.A. Bachman,
\emph{Relativistic Acoustic Doppler Effect in the Optical Limit},
\emph{Am. J. Phys.} {\bf 54} (1986) pg. 848-849

\bibitem{Eshelby}
J. D. Eshelby,
\emph{Uniformly Moving Dislocations},
\emph{Proc. Phys. Soc. A} {\bf 62} (1949) pg. 307-314

\bibitem{Bergman}
D.J. Bergman, E. Ben-Jacob, Y. Imry, K. Maki,
\emph{Sine Gordon solitons: Particles obeying relativistic dynamics},
\emph{Phys. Rev. A} {\bf 27} (1983) pg. 3345-3348

\bibitem{Unruh 1981}
W.G. Unruh,
\emph{Experimental Black-Hole Evaporation?},
\emph{Phys. Rev. Lett.} {\bf 46} (1981) pg. 1351-1353

\bibitem{Unruh 2014}
W.G. Unruh,
\emph{Has Hawking Radiation Been Measured?},
\emph{Foundations of Physics} {\bf 44} (2014)
doi:10.1007/s10701-014-9778-0

\bibitem{Gordon 1923}
Walter Gordon,
\emph{Zur Lichtfortpflanzung nach der Relativitatstheorie},
\emph{Annalen der Physik} {\bf 377} (1923) pg. 421-425
doi: 10.1002/andp.19233772202, consulted as “On the propagation of light in the theory of relativity”, translated by D. H. Delphenich
http://www.neo-classical-physics.info/uploads/3/4/3/6/34363841/gordon - optical metrics.pdf

\bibitem{Sfarti 2018}
A. Sfarti,
\emph{Optical Clock Behavior in a Gravitational Field},
\emph{Int. J. Photonics and Optical Technology} {\bf 4} (2018) pg. 20-23

\bibitem{Bravo 2023}
Tupac Bravo, Dennis Ratzel, and Ivette Fuentes,
\emph{Gravitational time dilation in extended quantum systems: the case of light clocks in Schwarzschild spacetime},
\emph{AVS Quantum Sci.} {\bf 5} (2023) pg. 014401
https://doi.org/10.1116/5.0123228

\bibitem{Ya’acov 2023}
Uri Ben-Ya’acov,
\emph{Time measurement with accelerating light-clocks},
\emph{Int. Assoc. for Relativistic Dynamics IARD-2022, Journal of Physics: Conference Series} {\bf 2482} (2023) pg. 012009
doi:10.1088/1742-6596/2482/1/012009

\bibitem{Wheeler 1955}
John Archibald Wheeler,
\emph{Geons},
\emph{Phys. Rev}{\bf 97} (1955) pg. 511-536

\bibitem{Brill and Hartle 1964}
Dieter R. Brill, James B. Hartle,
\emph{Method of the Self-Consistent Field in General Relativity and its Application to the Gravitational Geon},
\emph{Phys. Rev.} {\bf 135} (1964) pg. B271-B278

\bibitem{Guiot et al. 2020}
B. Guiot, A. Borquez, A. Deur and K. Werner,
\emph{Graviballs and Dark Matter},
\emph{JHEP} {\bf 11} (2020) pg. 159
doi:10.1007/JHEP11(2020)159

\bibitem{Delfino}
G. Delfino and G. Mussardo,
\emph{Non Integrable Aspects of the Multi-frequency
Sine Gordon Model},
\emph{International School for Advanced Studies, EP}{\bf 97} (1997)
arXiv:hep-th/9709028v1

\bibitem{CHOOZ 2012}
Double CHOOZ collaboration,
\emph{First test of Lorentz violation with a reactor-based antineutrino experiment},
\emph{Phys. Rev. D} {\bf 86} (2012) pg. 112009
arXiv:1209.5810

\bibitem{IceCube 2014}
Floyd W. Stecker,
\emph{Constraining Superluminal Electron and Neutrino Velocities using the 2010 Crab Nebula Flare and the IceCube PeV Neutrino Events},
\emph{Astroparticle Physics } {\bf 56} (2014) pg. 16-18
arXiv:1306.6095

\bibitem{D0 2015}
D0 Collaboration,
\emph{Search for Violation of CPT and Lorentz invariance in Bs meson oscillations},
\emph{Phys. Rev. Lett.} {\bf 115} (2015) pg. 161601
arXiv:1506.04123

\bibitem{New Era}
Gianfranco Bertone and Tim M.P. Tait,
\emph{A New Era in the Quest for Dark Matter},
\emph{Nature} {\bf 562} (2018) pg. 51-56
arXiv:1810.01668v1
doi:10.1038/s41586-018-0542-z

\bibitem{tachyons 2024}
Samuel H. Kramer and Ian H. Redmount,
\emph{Testing Tachyon-Dominated Cosmology with Type Ia Supernovae},
(2024)
arXiv:2403.13859

\bibitem{FeynmanLectures}
R.P. Feynman, R.B. Leighton, M. Sands,
\emph{The Feynman Lectures on Physics},
Vol. I, Ch. 16, Addison Wesley, 1963.

\bibitem{Roseveare 1982}
N.T. Roseveare,
\emph{Mercury's perihelion, from Le Verrier to Einstein},
Oxford University Press, New York, 1982. 

\bibitem{Lemmon and Mondragon 2016}
Tyler J. Lemmon and Antonio R. Mondragon,
\emph{Kepler's Orbits and Special Relativity In Introductory Classical Mechanics},
(2016)
https://doi.org/10.48550/arXiv.1012.5438 

\bibitem{lensing}
S.Vegetti, S. Birrer, G. Despali, C. D. Fassnacht, D. Gilman, Y. Hezaveh, L. Perreault Levasseur, J. P. McKean, D. M. Powell, C. M. O’Riordan, G.Vernardos,
\emph{Strong gravitational lensing as a probe of dark matter},
\emph{Space Sci. Rev.} {\bf 220} (2024)
arXiv:2306.11781v1 
https://doi.org/10.1007/s11214-024-01087-w

\bibitem{direct detection}
SuperCDMS Collaboration,
\emph{Status and prospects of the SuperCDMS Dark Matter experiment at SNOLAB},
(2024)
TAUP 2023, 076, published in PoS TAUP2023

\bibitem{54mps}
Cristian Armendariz-Picon and Jayanth T. Neelakanta,
\emph{How Cold is Cold Dark Matter?},
\emph{Journal of Cosmology and Astroparticle Physics} {\bf 2014} (2014)
arXiv:1309.6971v2
doi:10.1088/1475-7516/2014/03/049

\bibitem{metal poor}
Jonah Herzog-Arbeitman, Mariangela Lisanti, Lina Necib,
\emph{The Metal-Poor Stellar Halo in RAVE-TGAS and its Implications for the Velocity Distribution of Dark Matter},
\emph{Phys. Rev. Lett.} {\bf 120} (2017) pg. 041102
arXiv:1708.03635v2
doi:10.1088/1475-7516/2018/04/052

\end{thebibliography}
\end{document}